\documentclass[12pt]{article}

\usepackage{amsmath,amssymb,graphicx,cite} 
\usepackage{epsf}

\newcommand{\beq}{\begin{eqnarray}}
\newcommand{\eeq}{\end{eqnarray}}

\newcommand{\drawsquare}[2]{\hbox{%
\rule{#2pt}{#1pt}\hskip-#2pt
\rule{#1pt}{#2pt}\hskip-#1pt
\rule[#1pt]{#1pt}{#2pt}}\rule[#1pt]{#2pt}{#2pt}\hskip-#2pt
\rule{#2pt}{#1pt}}

\newcommand{\Yfund}{\drawsquare{7}{0.6}}
\newcommand{\Yasymm}{\drawsquare{7}{0.6}\hskip-7.6pt%
     \raisebox{7pt}{\drawsquare{7}{0.6}}}

\newcommand{\fund}{\drawsquare{6.5}{0.4}}

\newcommand{\centeron}[2]{{\setbox0=\hbox{#1}\setbox1=\hbox{#2}\ifdim
                            \wd1>\wd0\kern.5\wd1\kern-.5\wd0\fi \copy0
                            \kern-.5\wd0\kern-.5\wd1\copy1\ifdim\wd0>\wd1
                            \kern.5\wd0\kern-.5\wd1\fi}}
\newcommand{\ltap}{\>\centeron{\raise.35ex\hbox{$<$}}
                    {\lower.65ex\hbox{$\sim$}}\>}
\newcommand{\gtap}{\>\centeron{\raise.35ex\hbox{$>$}}
                    {\lower.65ex\hbox{$\sim$}}\>}

\newcommand\ZZ{\hbox{\zfont Z\kern-.4emZ}}
\font\zfont = cmss10 

\begin{document}
\begin{titlepage}
\begin{flushright}
{\tt  hep-th/0403062} \\
\end{flushright}

\vskip.5cm
\begin{center}
{\huge \bf  A Mixed Phase of SUSY Gauge
Theories from $a$-Maximization} \vskip.2cm
\end{center}

\begin{center}
{\bf {Csaba Cs\'aki}$^{a}$, {Patrick Meade}$^{a}$,
{and John Terning}$^{b}$} \\

\end{center}
\vskip 8pt

\begin{center}
  $^{a}${\it Institute for High Energy Phenomenology,
Laboratory of Elementary Particle Physics, \\ Cornell University,
Ithaca, NY 14853, USA } \\
\vspace*{0.1cm} $^{b}$ {\it Theory Division T-8, Los Alamos
National Laboratory, Los Alamos,
NM 87545, USA} \\
\vspace*{0.3cm}{\tt   csaki@mail.lepp.cornell.edu,
meade@mail.lepp.cornell.edu,terning@lanl.gov}
\end{center}

\vglue 0.3truecm

\begin{abstract}
\vskip 3pt \noindent

We study $\mathcal{N}=1$ supersymmetric $SU(N)$ gauge theories
with an antisymmetric tensor and $F$ flavors using the recent
proposal of $a$-maximization by Intriligator and Wecht.  This
theory had previously been studied using the method of
``deconfinement," but such an analysis was not conclusive since
anomalous dimensions in the non-perturbative regime could not be
calculated.  Using $a$-maximization we show that for a large range
of $F$ the theory is at an interacting superconformal fixed point.
However, we also find evidence that for a range of $F$ the theory
in the IR splits into a free ``magnetic" gauge sector and an
interacting superconformal sector.

\end{abstract}

\end{titlepage}

\newpage


\section{Introduction}
\label{sec:intro} \setcounter{equation}{0}
\setcounter{footnote}{0}

Finding the low-energy effective Lagrangian for a gauge theory in the strong
coupling regime is a virtually impossible task unless one appeals to lattice computations.
However, if one considers theories with a large symmetry group
then the symmetries could possibly be powerful enough to restrict
the structure of the low-energy effective Lagrangian. This is what
happens in supersymmetric~(SUSY) theories. Unbroken supersymmetry
implies that part (or all) of the Lagrangian is governed by
holomorphic objects, which  can sometimes be uniquely fixed based
on symmetry arguments and weak coupling limits. For
$\mathcal{N}=2$ theories the entire theory is determined by a
holomorphic quantity, the prepotential, which allows one to solve
the theory exactly~\cite{seibergwitten} in the IR.  In
$\mathcal{N}=1$ theories the entire Lagrangian is not governed by
a holomorphic object, but only the superpotential. The fact that
the superpotential is holomorphic in $\mathcal{N}=1$ theories
still allows one to make powerful statements about
non-perturbative physics and often lets one find the vacuum
structure and phase of the theory~\cite{seibergholo}. The possible phases for
theories with a small enough matter content are found to be
confining (with or without chiral symmetry breaking), a pure
abelian Coulomb phase (analogous to the ${\cal N}=2$ theories), or
the gauge group could be broken via a dynamically generated
superpotential. Theories with these phases have been completely
classified in~\cite{Seibergexact,sconfine,lowmatter}. Once the
matter content of a SUSY gauge theory is large enough, the known
possible phases are an interacting superconformal fixed point
(non-abelian Coulomb phase) or  free ``electric" or ``magnetic"
phases. The major tool used to study these phases is Seiberg
duality~\cite{seibergdual}, which is based on holomorphy and 't
Hooft anomaly matching. Seiberg duality is the IR equivalence of
two different SUSY gauge theories with the same flavor symmetries
and holomorphic invariants. This is often a strong-weak duality,
which means in the regime where one of the theories is strongly
coupled the dual is weakly coupled. The canonical examples of
Seiberg duality typically contain three regimes: where the
electric theory is IR free, while the magnetic theory is strongly
coupled (free electric theory), where both electric and magnetic
theories are interacting, but they correspond to the same IR fixed
point (non-abelian Coulomb phase), and where the dual is IR free
while the electric theory is strongly coupled (free magnetic
phase). However, there are relatively few examples of Seiberg
duality where the IR behavior is known, and thus one can only
guess the right low-energy description of most of the ${\cal N}=1$
theories.

An important step in finding a general prescription to determine
the low-energy description of most $\mathcal{N}=1$ gauge theories
has recently been made by Intriligator and Wecht~\cite{amax}. The
key ingredient is that when a SUSY theory is at a fixed point, it
necessarily has a larger space-time symmetry group, the
superconformal group.  A particular $U(1)_R$ symmetry plays a
special role since its corresponding R-charge  is one of the
generators of the superconformal group. From the superconformal
algebra it follows \cite{Flato:1983te,Dobrev:qv} that the R-charge
and  the dimension $\Delta$ of a chiral operator $\mathcal{O}$
satisfy
\begin{equation}
\frac{3}{2}R(\mathcal{O})=\Delta(\mathcal{O}).
\end{equation}
Therefore we can determine the anomalous dimensions of the entire
chiral ring from the fact that $\Delta\equiv 1+\gamma/2$ if we can
determine their $R$-charges.  Thus we can ``solve" the gauge
theory if we can determine the $R$-symmetry $U(1)_R\subset
SU(2,2\vert 1)$ of the superconformal algebra.  This is an easy
problem in SUSY QCD where there is no ambiguity in determining the
$U(1)_R$ symmetry. However when there are additional fields there
are additional $U(1)$ flavor symmetries and and one can form a
linear combination of these anomaly free $U(1)$'s. It is then not
clear what principle will determine which of these linear
combinations will be the preferred $U(1)_R$ that appears in the
superconformal algebra. This is the problem that has been recently
solved by Intriligator and Wecht~\cite{amax} through a process
called $a$-maximization. They found that the $R$-symmetry
appearing in the superconformal algebra is the one that maximizes
a central charge called $a$. A brief review of this process will
be presented at the beginning of the next section.

\par The ability to find the superconformal $U(1)_R$ symmetry is a
major step forward in exact results in SUSY gauge theories.
However, finding the superconformal $U(1)_R$ still does not solve
the entire theory since there is no way from $a$-maximization to
determine where the superconformal phase ends.  The process of
$a$-maximization has only been implemented in a few cases such as
for $SU(N)$ gauge theories with one and two adjoints and various
superpotential perturbations~\cite{amax,kutasov,int2}, all of
which are vector-like theories (it has also been examined in the
case of general theories in a different
framework~\cite{kutasov2}). However, some of the most interesting
SUSY gauge theories are the chiral theories, since these are the
ones that can lead to dynamical supersymmetry breaking. The
simplest models of dynamical SUSY breaking usually utilize a gauge
theory with an antisymmetric tensor and some number of
flavors~\cite{ADS,32,otherdsb}. Therefore, a lot of effort was
expended during the nineties to try to understand the dynamics of
such
theories~\cite{hitoshi,erich,berkooz,pouliot,cascade,fiveeasy}.
For a small number of flavors the dynamics of the theory is
well-understood~\cite{erich,pouliot,sconfine}, however one could
not conclusively find the low-energy phase of such a theory for an
arbitrary number of flavors.

In light of the new developments we return in this paper to the
study of the dynamics of a supersymmetric $SU(N)$ gauge theory
with one two-index antisymmetric tensor, $F$ fundamentals, and
$N+F-4$ antifundamentals and no tree-level superpotential. The new
methods will allow us to finally pin down the phase structure of
this model. In studying this theory we implement $a$-maximization
first directly, and then consider a dual of this theory based on
the method of ``deconfinement"~\cite{berkooz}. The process of
``deconfinement" allows one to come up with another strongly
coupled description of the original theory using only fundamental
fields which can then be dualized using ordinary Seiberg duality.
Studying this deconfined dual will let us explore what happens to
the theory when some of the gauge invariants in the electric
theory go free.  The dual of this theory using ``deconfinement"
has been previously studied in~\cite{pouliot,fiveeasy}.  However,
since the superconformal R-symmetry could not be found before the
method of $a$-maximization was known, one could not draw definite
conclusions about the phase of the theory even using the
deconfined dual.

One of the advantages of considering the deconfined dual in the
case of the antisymmetric tensor is that the fields that go free
as the number of flavors is reduced will be elementary fields in
the dual theory.  Therefore the procedure suggested
in~\cite{kutasov} of eliminating the contributions of the free
field from $a$ in the $a$-maximization procedure is
straightforward to carry out.   One can also check explicitly if a
field going free would also imply the existence of a new phase or
not. The deconfined dual has a product group structure.
In~\cite{fiveeasy} hints were found  that for $F=5$ and for $N>6$
a new type of mixed phase occurs where one of the gauge groups
remains at a superconformal fixed point whereas the other group
becomes IR free.  In this paper we use $a$-maximization in
combination with ``deconfinement" and find strong evidence for the
fact that this new mixed phase exists and determine exactly when
it occurs. What this implies is that the original electric theory
after a certain point ceases to be a good description of the
physics and actually splits into two sectors:
an interacting non-abelian Coulomb phase and a co-existing free magnetic
phase. The structure of the paper is as follows: in
Section~\ref{selectric} we outline the original theory and give
the results of $a$-maximization for this ``electric" theory.
In Section~\ref{deconfines} we explain
the method of ``deconfinement" and show how to use it to find a
dual description of the electric theory.  In
Section~\ref{newphase} we then implement $a$-maximization in the
deconfined dual theory, show explicitly how the decoupling of the
free fields happens in the dual description, and discuss the
arguments for the appearance of a mixed phase in this theory.

\section{$a$-maximization in the electric theory}
\label{selectric} \setcounter{equation}{0}
\par

Before we start analyzing the $SU(N)$ theory with an antisymmetric
tensor, we will briefly review the central charge $a$ and
$a$-maximization for those not familiar with the original paper of
Intriligator and Wecht~\cite{amax}.  In a SUSY gauge theory the
trace anomaly of the stress-energy tensor, $T^{\mu\nu}$, has both
internal contributions (from the gauge sector) and external
contributions from external background sources that are coupled to
currents in the theory. The central charge $a$ of a
four-dimensional superconformal gauge theory is the coefficient of
the contribution from an external supergravity background.  The
definition of $a$ comes from coupling the stress energy tensor to
a background metric $g_{\mu\nu}(x)$ which then shows up in the
trace anomaly as
\begin{equation}
T^\mu_\mu=\Theta \sim
\frac{1}{g^3}\tilde{\beta}(F_{\mu\nu}^a)^2-
a(g)(R_{\mu\nu\rho\sigma})^2+\dots
\end{equation}
where $g$ is the gauge coupling, $\tilde{\beta}$ is the numerator
of the exact NSVZ $\beta$ function~\cite{NSVZ}, $F_{\mu\nu}^a$ is
the gauge field strength,  and $R$ is the curvature tensor whose
square is the Euler density.  The central charge $a$ was
conjectured by Cardy~\cite{Cardy} to satisfy a four dimensional
version of the Zamolodchikov $c$-theorem~\cite{ctheorem}: $a_{IR}<
a_{UV}$. The connection between the $U(1)_R$ symmetry that is in
the superconformal algebra and $a$ is that $a$ can be expressed in
terms of 't Hooft anomalies of this particular
$R$-symmetry~\cite{anselmi1,anselmi2}.  The relation between $a$
and the 't Hooft anomalies is
\begin{equation}\label{arsym}
a=\frac{3}{32}\left(3 \mathrm{Tr }R^3-\mathrm{Tr }R \right).
\end{equation}
This relation still does not tell us what the superconformal
$U(1)_R$ symmetry is since neither side of Eq.~(\ref{arsym}) is
fixed at this point.

Let us consider a trial $R$-symmetry made up of some arbitrarily
chosen initial $R$-symmetry $R_0$ and the various additional
$U(1)$ symmetries $Q_I$ of the global symmetry group of the theory
\begin{equation}\label{rtrial}
R_{trial}=R_0+\sum_i s_I Q_I
\end{equation}
where $s_I$ are arbitrary real coefficients that tell us the
admixture of symmetries making up our trial $R$-symmetry.  What
Intriligator and Wecht have shown is that the $s_I$ corresponding
to the linear combination that gives the superconformal
$R$-symmetry, $\hat{s}_I$, come from maximizing the central charge
$a_{trial}$. $a_{trial}$ is constructed by using the trial
$R$-symmetry in Eq.~(\ref{rtrial}) with Eq.~(\ref{arsym}).  The
condition that $a$ is maximized implies that the first derivatives
of $a_{trial}$ with respect to the $s_I$ vanish which implies
\begin{equation}
\frac{\partial a_{trial}}{\partial
s_I}=\frac{3}{32}\left(9\mathrm{Tr}R_{trial}^2
Q_I-\mathrm{Tr}Q_I\right)=0.
\end{equation}
Thus the first condition of a-maximization is
\begin{equation}\label{quadeq}
9\mathrm{Tr}(R^2 Q_I)=\mathrm{Tr}Q_I.
\end{equation}
To find a local maximum the second condition is that the matrix of
second derivatives
\begin{equation}\label{negdef}
\frac{\partial^2 a_{trial}}{\partial s_I\partial s_J
}=\frac{27}{16}\mathrm{Tr}R_{trial} Q_I Q_J < 0
\end{equation}
is negative-definite.  Intriligator and Wecht showed
in~\cite{amax} that Eqs.~(\ref{quadeq}) and (\ref{negdef}) were
always true for any unitary superconformal field theory. Therefore
maximizing $a$ over the space of possible $R$-symmetries
determines the superconformal $U(1)_R$.

The process of $a$-maximization relies on being able to identify
the superconformal $R$-symmetry from the weakly coupled UV fixed
point, assuming that the global symmetries of the IR
superconformal theory match those of the UV theory. However, in
many cases there are accidental symmetries in the IR, which can
also be part of the superconformal $R$-symmetry. For instance such
an accidental symmetry appears when one of the gauge invariants
becomes a free field. At such points $a$-maximization could
conceivably break down, since the theory will not necessarily
remain in the superconformal phase. For example a field going free
could signal the appearance of a free magnetic phase. The only way
to unambiguously decide what is exactly happening at such points
is if one has a weakly coupled dual at hand, which is usually not
the case.

\par

We are interested in the low-energy behavior of a supersymmetric
$SU(N)$ gauge theory with one two-index antisymmetric tensor, $F$
fundamentals  and $F+N-4$ antifundamentals.  We would like to use
the method of $a$-maximization which is applicable in the
superconformal phase. Thus we first look for the Banks-Zaks fixed
point\cite{bankszaks} which occurs for $F=2N-3-\epsilon N$.  To
simplify the expressions we will work explicitly in the large
$N$,$F$ limit with $x\equiv N/F$ held fixed.  The theory has the
following transformation properties under the non-abelian flavor
symmetries and the superconformal $R$-symmetry:

\beq \label{1asym}
\begin{tabular}{c|c|c|c|c}
  & $SU(N)$ & $SU(F)$  & $SU(F+N-4)$ & $U(1)_R$ \\
\hline $Q$ & \fund & $\fund$ & {\bf 1} & $R(Q)$ \\
$\overline{Q}$ & $\overline{\fund}$   & {\bf 1} & \fund &
$R(\overline{Q})$ \\
${\rm A}$ & \Yasymm & {\bf 1} & {\bf 1} & $R({\rm A})$\\
\end{tabular}
\eeq

The vanishing of the NSVZ $\beta$ function is necessary for the
theory to be at a superconformal fixed point. This condition is
equivalent to the cancellation of the $SU(N)^2 U(1)_R$ anomaly for
the superconformal $R$-symmetry, and implies that
\begin{equation}
R({\rm A}) = \frac{2-R(Q)-(x+1) R(\overline{Q})}{x}.
\end{equation}

Applying the $a$-maximization procedure we find that
\beq
R(Q)=
R(\overline{Q})=-\frac{12-9x^2+\sqrt{x^2(-4+x(73x-4))}}{3(-4+(x-4)x)}.
\eeq

\begin{figure}[ht]
\centerline{\includegraphics[width=0.65\hsize]{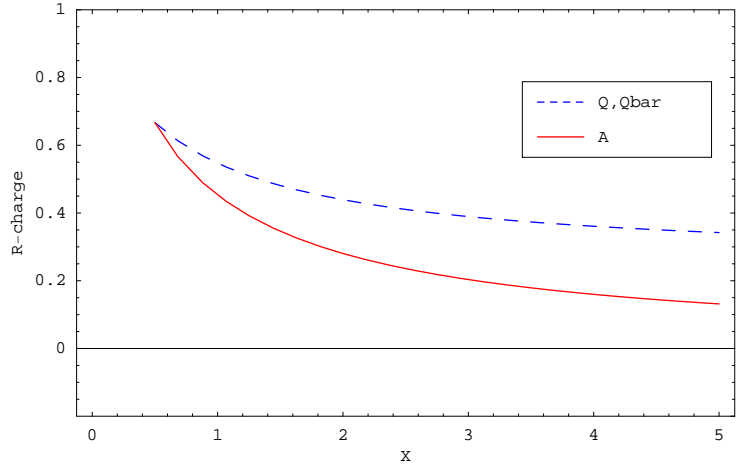}}
\caption{The naive $R$-charges of the fields are plotted as a
function of $x$, the dashed line represents the $R$-charges of
$Q$,$\overline{Q}$ and the solid line is the anti-symmetric tensor
$A$. For $x>x_M\sim 2.95367$ the $R$-charges will be modified when
taking into account unitarity constraints.} \label{1aplot}
\end{figure}

The flow of the $R$-charges as a function of $x$ is shown in
Fig.~\ref{1aplot} which starts at a value of $x=.5$, corresponding
to the Banks-Zaks fixed point in the large $N$,$F$ limit.  One
should note that the $R$-charges of $Q$ and $\overline{Q}$ are the
same even though the theory is chiral. This corresponds to the
fact that to all orders in perturbation theory the anomalous
dimensions for $Q$ and $\overline{Q}$ have to agree in the absence
of a superpotential, since gauge interactions do not distinguish
between the two fields.  The $R$-charges in Fig.~\ref{1aplot} will
be modified for $x>x_M\sim2.95367$ due to unitarity constraints
that we will discuss.

\par The chiral ring of this theory is made up of two types of
mesons, $M=Q\overline{Q}$ and $H=\overline{Q}A\overline{Q}$, as
well as baryons of the form $B_k= Q^k A^{\frac{N-k}{2}}$ (for
$k$,$N$ both even or odd and $k \le min(N,F)$) and
$B=\overline{Q}^N$. Unitarity constrains the dimensions of the
operators in the chiral ring to be greater than one. A possible
signal for a theory to leave the superconformal phase is when
there is an apparent violation of the unitarity constraint (for
example in SUSY QCD, the meson becoming a free field signals the
onset of the free magnetic phase).  In the theory with the
antisymmetric tensor under consideration here the smallest
invariant (in terms of number of fields) is $M$, therefore this
field is likely to go free first as $x$ is increased which is what
we find. We find the point at which the meson becomes a free field
is at
\begin{equation}\label{mfree}
x=x_M=\frac{4}{9}\left(4+\sqrt{7}\right)\sim 2.95367.
\end{equation}
There are then two possibilities: either the theory is out of the
superconformal phase and $a$-maximization should no longer be used, or
it is
also possible that at the point where $M$
first appears to violate the unitarity bound the meson becomes a
free field while the other members of the chiral ring are still
interacting. It is impossible to decide just based on the electric
theory
which of these possibilities actually occurs, but for now we will
assume that
it is the latter case (that is $M$ becomes free while the other fields
remain interacting). We will see more compelling evidence for this from
the
deconfined dual description in the the next section.

\begin{figure}[ht]
\centerline{\includegraphics[width=0.65\hsize]{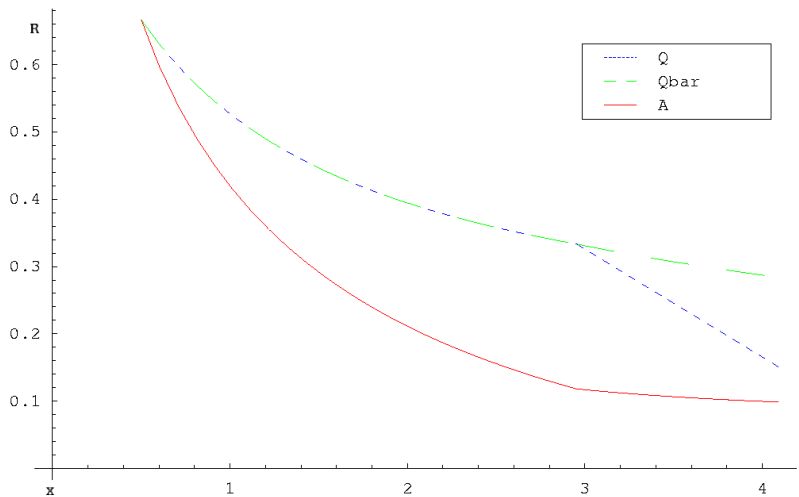}}
\caption{The $R$-charges of the fields are plotted as a function
of $x$ taking into account that $M$ went free.  The short dashed
line represents the $R$-charges of $Q$, and the long dashed line
represents $\overline{Q}$ while the solid line is the
anti-symmetric tensor $A$.
} \label{rchargemfree}
\end{figure}

If the meson becomes a free field, there will be an additional
$U(1)$ symmetry not present in the UV description which can mix
into the superconformal $R$-symmetry and which will ensure that
the R-charge of $M$ is 2/3. One then subtracts the contribution of
the meson from the $a$ used originally and re-maximizes to find
the flow for values of $x$ greater than those in Eq.~(\ref{mfree})
as first described in~\cite{kutasov}.  Following the procedure set
out in~\cite{kutasov} once the meson $M$ goes free we construct a
new $a_{int}$ made up of only the assumed interacting sector of
the theory \beq a_{int}&=&a_0-a(R(M))
\\
a_{int}&=&a_0-\frac{3}{32} F (F+N-4)\big(3
(R(Q)+R(\overline{Q})-1)^3-(R(Q)+R(\overline{Q})-1)),
\eeq
where $a_0$ is the original $a$ of the theory.  Maximizing
$a_{int}$ now determines the $R$-charges of $Q$,$\overline{Q}$ and
$A$ for $x>x_M$.  Because the
analytical expressions for the $R$-charges become too complicated
to present here we simply plot the results in
Fig.~\ref{rchargemfree}.

\par

The second meson $H=\overline{Q}A\overline{Q}$ also goes free when
$x=x_H\sim 4.08952$.  Therefore in Fig.~\ref{rchargemfree} for
$x>x_H$ the R-charges will be modified when taking into account
unitarity constraints.  That $H$ goes free at this value of $x$
depends upon the large $N$ limit, for $N<8$ $H$ will not become a
free field before the theory confines for $F<5$. When $H$ goes
free one might expect that the same thing happens as when $M$ went
free, i.e. it decouples and the rest of the theory remains
interacting.  We do not continue this process in this ``electric"
description because in a dual ``deconfined" description we believe
$H$ going free signals a change from a purely interacting
non-abelian Coulomb phase into a mixed phase that we will describe
in Section~\ref{newphase}.

\section{Deconfinement}\label{deconfines}

No weakly coupled dual with one gauge group is known for the
theory under consideration.  This is unfortunate since as was
pointed out in~\cite{kutasov} one does not know when
$a$-maximization breaks down and a free magnetic phase occurs
without a Seiberg dual. There is the possibility though that there
is another strongly coupled description of the same physics.  We
can  find such a strongly coupled description of the $SU(N)$ gauge
theory in question if we view the antisymmetric tensor as a
composite coming from an s-confining (confining without chiral
symmetry breaking and with a confining superpotential) $Sp$ gauge group: this idea
is known as ``deconfinement" \cite{berkooz}.  It is a rather
straightforward process since s-confining SUSY gauge theories are
well documented and classified~\cite{sconfine}.  Once one has a
new strongly coupled description of the physics in terms of only
fundamental representations of $SU(N)$ one applies the usual
Seiberg dualities to find further dual descriptions of the
physics.  As we will see after applying Seiberg duality, instead
of having two groups where one is in a non-abelian Coulomb phase
and the other is confining, both will be in a non-abelian coulomb
phase for small enough $x$.  With that in mind let us now look at
the details of this procedure for our theory in question which has
been examined in \cite{pouliot,fiveeasy}.  One can skip ahead to
the final (second) dual description since that is all we will use here but
we have included the intermediate steps for completeness.

We start with the $SU(N)$ gauge theory with an antisymmetric
tensor \cite{pouliot,fiveeasy} for odd $N$ and with $F\ge 5$
flavors:
\begin{equation}
\label{elec}
\begin{tabular}{c|c|cc}
  & $SU(N)$ & $SU(F)$  & $SU(F+N-4)$\\
\hline $Q$ & \fund & $\fund$ & {\bf 1}\\
$\overline{Q}$ & $\overline{\fund}$   & {\bf 1} & \fund \\
${\rm A}$ & \Yasymm & {\bf 1} & {\bf 1}
\end{tabular}.
\end{equation}

Deconfinement means that instead of considering the above theory with the antisymmetric
tensor, one imagines that this antisymmetric tensor is a composite meson of
another strongly interacting gauge group that confined before the $SU(N)$ group
became strongly interacting. Thus we assume that there is a gauge group $G$ which has a weakly
gauged flavor symmetry $SU(N)$. Since we want the meson to be in an antisymmetric
representation of this $SU(N)$, the gauge group should be chosen to be an $Sp$ group.
It is well-known that $Sp(M)$ is s-confining (that is confining without chiral symmetry
breaking) if the number of fundamentals under $Sp(M)$ are $F=M+4$. However, this confining
group will generate a superpotential for the confined meson. In order to eliminate this superpotential
{\it after} confinement one needs to add an additional
 superpotential term to the theory {\it before} confinement.
The role of this term in the superpotential will be to set some fields to zero after confinement and
thereby eliminating the entire confining superpotential. The details of this procedure for deconfining
arbitrary two-index representation are described in~\cite{cascade}. Here we just repeat the main steps
leading to the final dual description that we will be using.

The deconfined dual description of this theory is then found by taking
$A$ to be a composite meson of an s-confining $Sp(N-3)$ theory \beq
\begin{array}{c|cc|ccccc}
  & SU(N)  & Sp(N-3)  & SU(F)  & SU(N+F-4)
\\
\hline
Y & \Yfund & \Yfund & {\bf 1} & {\bf 1}  \\
Z & {\bf 1} & \Yfund & {\bf 1} & {\bf 1} \\
\overline{P} & \overline{\Yfund} & {\bf 1} & {\bf 1} & {\bf 1} \\
Q & \Yfund & {\bf 1} & \Yfund & {\bf 1}  \\
\overline{ Q}  & \overline{\Yfund} & {\bf 1} & {\bf 1}  & \Yfund \\
\end{array}
\eeq
with a superpotential
\beq
W=Y Z \overline{P}~.
\eeq
When the $Sp(N-3)$ group confines, the superpotential becomes a mass term for one of the meson components,
which eliminates the entire confining superpotential.

The $SU(N)$ group has $N+F-3$ flavors so we can use Seiberg
duality for $SU$ groups to find a dual for this
deconfined theory: \beq
\begin{array}{c|cc|cc}
  & SU(F-3)  & Sp(N-3)  & SU(F)  & SU(N+F-4)
\\
\hline
y & \Yfund & \Yfund & {\bf 1} & {\bf 1}  \\
\overline{p} & \overline{\Yfund} & {\bf 1} & {\bf 1} & {\bf 1}  \\
q & \Yfund & {\bf 1} & \overline{\Yfund} & {\bf 1}   \\
\overline{ q} & \overline{\Yfund} & {\bf 1}  & {\bf 1}  &
\overline{\Yfund}  \\
M & {\bf 1} & {\bf 1} & \Yfund & \Yfund \\
L & {\bf 1} & \Yfund & {\bf 1} & \Yfund \\
B_1 & {\bf 1} & {\bf 1} & \Yfund & {\bf 1} \\
\end{array}
\eeq with \beq W= M q \overline{q} + B_1 q \overline{p} +L y
\overline{q}~.
\eeq

In this first dual we have an $Sp(N-3)$ group with $N+2F-7$ fundamentals. We can then use
Seiberg duality for $Sp$ groups~\cite{Spduality} to find
an $Sp(2F-8)$ dual, which after integrating out massive fields
has the following content

\beq \label{finaldual}
\begin{array}{c|cc|cc}
  & SU(F-3)  & Sp(2F-8)  & SU(F)  & SU(N+F-4)
\\
\hline
\tilde{y} & \overline{\Yfund} & \Yfund & {\bf 1} & {\bf 1}  \\
\overline{p} & \overline{\Yfund} & {\bf 1} & {\bf 1} & {\bf 1}  \\
q &\Yfund  & {\bf 1} & \overline{\Yfund} & {\bf 1}   \\
M & {\bf 1} & {\bf 1} & \Yfund & \Yfund \\
l & {\bf 1} & \Yfund & {\bf 1} & \overline{\Yfund}\\
B_1 & {\bf 1} & {\bf 1} & \Yfund & {\bf 1} \\
a & \Yasymm & {\bf 1} & {\bf 1} & {\bf 1} \\
H & {\bf 1} & {\bf 1} & {\bf 1} & \Yasymm  \\
\end{array}
\eeq

and superpotential
  \beq \label{dualw} W=M q l \tilde y + H
ll+ B_1 q \overline{p}+ a \tilde{y} \tilde{y} ~. \eeq In the final
dual description we see that both $M=Q\overline{Q}$ and
$H=\overline{Q}A\overline{Q}$ are mapped to fundamental fields.
This is important, because in the process of $a$-maximization it
is easier to decide what happens to the theory when fundamental
fields rather than composite objects become free. To even have the
possibility of a weakly coupled description, a field that becomes
free must be fundamental in that description.

One can similarly find a dual for the case when $N$ is even by following the same steps
outlined above. The final dual will be:

\beq \label{finaldualNeven}
\begin{array}{c|cc|ccc}
  & SU(F-2)  & Sp(2F-8)  & SU(F)  & SU(N+F-4) & SU(2)
\\
\hline
\tilde{y} & \overline{\Yfund} & \Yfund & {\bf 1} & {\bf 1} & {\bf 1} \\
\overline{p} & \overline{\Yfund} & {\bf 1} & {\bf 1} & {\bf 1} & \Yfund \\
q &\Yfund  & {\bf 1} & \overline{\Yfund} & {\bf 1} & {\bf 1}  \\
M & {\bf 1} & {\bf 1} & \Yfund & \Yfund & {\bf 1}\\
l & {\bf 1} & \Yfund & {\bf 1} & \overline{\Yfund}& {\bf 1}\\
S & {\bf 1} & {\bf 1} & \Yfund & {\bf 1}& \Yfund \\
a & \Yasymm & {\bf 1} & {\bf 1} & {\bf 1}& {\bf 1} \\
H & {\bf 1} & {\bf 1} & {\bf 1} & \Yasymm  & {\bf 1}\\
B_0 & {\bf 1} & {\bf 1} & {\bf 1} & {\bf 1} & {\bf 1}\\
\end{array}
\eeq with the superpotential \beq \label{dualwNeven} W=M q l
\tilde y + H ll+ S q \overline{p}+ a \tilde{y} \tilde{y} +B_0 a
\overline{p}^2~. \eeq Note the appearance of the additional
(spurious) $SU(2)$ global symmetry. This extra $SU(2)$ is
necessary to avoid the appearance of a Witten-anomaly in the Sp
group. Note however, that none of the fields that transform under
the $Sp(2F-8)$ group transform under this $SU(2)$. We should
remark that there are many other possible ways to deconfine the
antisymmetric tensor
 both for $N$ odd and $N$ even, in which case the analog of the spurious global symmetry
would be bigger, an $SU(K)$ group. However, in the second dual one always ends up with
an $Sp(2F-8)$ group with the same matter content and none of these fields transforming under
the $Sp(2F-8)$ group will transform under the $SU(K)$.

\section{$a$-maximization in the deconfined dual theory: evidence for a
mixed phase} \label{newphase} \setcounter{equation}{0}

In the final dual description, Eq.~(\ref{finaldual}), given that
the mesons $M$ and $H$ are fundamental fields, their interactions
are only dependent upon the couplings in the superpotential which
contribute to their anomalous dimensions.  The hypothesis in the
electric description was that when $M$ and $H$ go free that they
simply decouple while the rest of the theory remains interacting.
Here we show evidence that this is in fact the case when $M$ goes
free, but not when the $H$ meson goes free.
\par
The deconfined dual description in Eq.~(\ref{finaldual}) is a
strongly coupled description of the electric theory. In the
electric picture there exists a weakly coupled Banks-Zaks fixed
point~\cite{bankszaks}, however in the deconfined dual description
there is no weakly coupled fixed point due to the product group
structure.  Therefore one doesn't really know if the terms in the
deconfined superpotential~(\ref{dualw}) are actually relevant or
not.  The best one can do is to assume the superpotential is
relevant and thus deconfinement is valid around the Banks-Zaks
fixed point of the electric theory. Then one can apply the
$a$-maximization process to the final dual description to find the
dimensions of the fields.  By construction one then finds that the dimensions of the
fields in the chiral ring of the ``electric" theory agree with
those of the final dual near the Banks-Zaks fixed point of the
original theory.
In this final dual description we find that $M$ goes free at
$x=x_M$ which corresponds identically to the electric description.
When $M$ goes free in the magnetic description the superpotential
coupling in front of $M q l \tilde y$ must vanish. However, this
does not imply any further field necessarily going free, and it is
consistent to assume that due to this coupling flowing to zero $M$
becomes a free field, while all other fields remain interacting.
This is a very explicit realization of the procedure
of~\cite{kutasov} about incorporating accidental symmetries into
$a$-maximization.

Thus it is reasonable to continue
on with the assumption that when $M$ goes free it simply decouples
and we then find that $H$ goes free at $x=x_H$, again exactly at
the same point where it happens in the electric theory.
However, the fact that $H$ becomes a free field also implies that
at this point its superpotential coupling flows to zero. Assuming
continuity of the anomalous dimension in $x$ we can also deduce
that the dimension of the dual quark, $\Delta (l)$, is equal to
one, since the superpotential term $Hll$ has $R$-charge 2. The
anomalous dimension of $l$ can be expressed as $\gamma_l=g_{Sp}^2
h(\mathrm{couplings})$ where $h$ is a function of possibly all the
couplings in the theory.  The fact that the anomalous dimension of
$l$ vanishes at the point where $H$ goes free could either be a
consequence of an unexpected cancellation which causes there to be
a zero of $h$ when $H$ goes free, or more simply it could imply
that the gauge coupling of the $Sp$ group also vanishes at this
point.  In the latter case $l$ would also become a free field
implying $\gamma_l=0$ as required, but this also implies that the
whole $Sp$ group becomes free at the point where $H$ goes free.
This would be analogous to the case of SUSY QCD where the meson
going free implied the whole dual gauge group became free, i.e.
the theory entered a free magnetic phase. We find it more
plausible that the $Sp$ gauge group becomes free when $H$ goes
free. This is also the scenario that one would expect to happen
based on estimates of the value of the $\beta$-function for the
$Sp$ group, which supports the claim that the $Sp$ group gauge
coupling in the IR indeed vanishes and $Sp$ is IR free for
sufficiently large $x$.
\par
Next we will show how to actually estimate the value of the
$\beta$-function for the $Sp$ group. This will be similar to the
argument found for the special case of five flavors in
ref.~\cite{fiveeasy}. We can simplify the analysis by noting that
the ratio of the two holomorphic scales in the the final dual
description $\Lambda_{Sp}$ and $\Lambda_{SU}$ can be varied
arbitrarily. Because of holomorphy we know that there can be no
phase transition as the ratio is varied, thus we can always go to
a limit where one of the gauge couplings is as small as we desire.
Thus we will work in the limit $\Lambda_{Sp} \ll \Lambda_{SU}$ and
show that the $\beta$ function of the $Sp$ group is positive for
large enough $x$, which will thus imply that the $Sp$ group is
indeed IR free. The bound on $x$ that we find below is in
accordance with the exact value found from $a$-maximization.

In the
limit where $g_{Sp}$ goes to zero we can expand the $\beta$ function of the $Sp$ group
perturbatively in the small coupling limit:
\begin{equation}\label{beta}
\beta(g_{Sp})=-\frac{g_{Sp}^3}{16\pi^2}\left[3(2F-6)-(F-3)(1-
\gamma_{\tilde{y}}(g_{Sp}=0))-(N+F-4)\right]+\mathcal{O}(g_{Sp}^5).
\end{equation}
Here we have used the fact that $\gamma_{l}(g_{Sp}=0)=0$. This follows from the
fact that in the limit $g_{Sp}\to 0$ $l$ is  a  gauge invariant, and it should
obey $\Delta_l\ge 1$. However, it appears in a superpotential term
$Hll$, where all three fields are gauge invariant in the zero $Sp$ coupling limit.
Since all of these fields have at least dimension one, this term has to
be irrelevant, and so in this limit $l$ is a free field, implying
 $\gamma_{l}(g_{Sp}=0)=0$.

We therefore see from (\ref{beta}) that if
\begin{equation}\label{irfree}
4F-N-11+(F-3)\gamma_{\tilde{y}} \le 0
\end{equation}
the $Sp$ group is IR free.  The only thing we need to find is a bound on the
anomalous dimension of $\tilde{y}$.
We can get a bound on $\gamma_{\tilde{y}}$ in the following way:
if we look at the last term in the superpotential (\ref{dualw}) we
see that the coefficient of this term by assumption does not
vanish in the limit $g_{Sp}$ goes to zero. Therefore the
$R$-charge in this superpotential must add up to two. Assuming
that the $SU$ group is at a conformal fixed point we can use the
relation between $R$-charges and anomalous dimensions to find
\begin{equation}\label{anom}
\gamma_{a}+2\gamma_{\tilde{y}}=0.
\end{equation}

Thus we can get an upper bound on $\gamma_{\tilde{y}}$ if we find a lower bound on $\gamma_{a}$.
Such a bound can be obtained by considering the unitarity bound for a gauge invariant
containing only $a$'s.
Let us first suppose that $F$ is odd, in this case $a^{\frac{F-3}{2}}$
is such a gauge invariant and maps to the field $B_F=Q^F
A^{\frac{N-F}{2}}$ of the ``electric" theory.  The reason why we emphasize this is because the
unitarity bound should only be imposed on fields that are part of the chiral ring. There do exist
gauge invariant operators (for example $q\overline{p}$) that are lifted by the F-flatness conditions
and therefore the unitarity bound should not be imposed on them.
Since $B_F=Q^F
A^{\frac{N-F}{2}}$ is a gauge
invariant of the electric theory it is bounded by the unitarity
constraint to have
\begin{equation}\label{bound}
\frac{F-3}{2}+\frac{F-3}{4}\gamma_{a}\ge 1,
\end{equation}
since $\Delta_i=1+\gamma_i/2$.  By combing Eq.~(\ref{anom}) and
Eq.~(\ref{bound}) and taking the large $F$ limit we see that
\begin{equation}\label{boundy}
\gamma_{\tilde{y}} \le 1.
\end{equation}
If we use this bound for $\gamma_{\tilde{y}}$ and combine it with
Eq.~(\ref{irfree}) in the large $N,F$ limit we see that for
\begin{equation}
N\ge 5F
\end{equation}
or
\begin{equation}\label{irbound}
x\ge 5
\end{equation}
the $Sp$ gauge group is IR free.  This is consistent with the
result obtained from $a$-maximization, and gives strong support
for the expectation that for $x>x_H$ the $Sp$ group is indeed IR
free.  One should note that this bound is different than the naive
estimate of where the $Sp$ group would go free if you did not take
into account the dynamics of the $SU$ group.  Ignoring the $SU$
dynamics one would find that for $x \ge 4$ the $Sp$ group was IR
free but as we see including the effect of the superpotential and
the $SU$ strong dynamics we find a higher value of $x$ at which
$Sp$ becomes free which is consistent with $a$-maximization.

For the case that $F$ is even we
will have to use a different gauge invariant to find a bound on
$\gamma_{\tilde{y}}$.  In this case we will use the invariant $q
a^\frac{F-4}{2}$ which maps to $B_{F-1}=Q^{F-1}
A^{\frac{N-F-1}{2}}$ of the electric theory which gives a
unitarity constraint of
\begin{equation}\label{boundeven}
F-4+\gamma_q+\frac{F-4}{2}\gamma_{a}\ge 0.
\end{equation}
Since Eq.~(\ref{anom}) does not depend on whether $F$ is even nor
odd, we can use Eq.~(\ref{anom}) with Eq.~(\ref{boundeven}) to
show that
\begin{equation}\label{almostbound}
F-4+\gamma_q\ge (F-4)\gamma_{\tilde{y}}.
\end{equation}
Using the $a$-maximization process we find that $\gamma_q$ is
bounded for all values of $x$ therefore in the large $F$ limit
Eq.~(\ref{almostbound}) reduces to Eq.~(\ref{boundy}).  We
therefore find that for $F$ even or odd for $x>5$ the $Sp$ group
becomes IR free. To summarize, we have found that there is a value
of $x$ above which the $Sp$ gauge group is necessarily IR free.
Since the anomalous dimension of $l$ vanishes when $H$ goes free,
it is very plausible to identify the onset of the IR free phase of
the $Sp$ group with the point when $H$ goes free, as argued when
discussing the results of $a$-maximization. For the case when $N$
is even one needs to use the dual in (\ref{finaldualNeven}). We
find identical results in this case in the large, $N,F$ limit: the
meson $M$ goes free at $x_M$, $H$ goes free at $x_H$ and at that
point the dual $Sp(2F-8)$ group becomes IR free. One may worry
about the appearance of a different $SU$ gauge group and the
additional $SU(2)$ global symmetry in the final dual in this case.
This is however not a real concern, since the fields that we claim
to be free do not transform under the $SU$ gauge symmetry nor
under the extra global $SU(2)$. One can also check, that our
results about the free dual quarks and gauge fields are
independent of the particular choice of the deconfining gauge
group: the second dual will always have the same $Sp(2F-8)$ factor
with the same degrees of freedom transforming under it, which has
to be the case if the $Sp$ group is indeed IR free.

Thus what we have found is that there is a phase in the theory
with the antisymmetric tensor which is similar to the free
magnetic phase of SUSY QCD, since there is an IR free gauge group
with free magnetic gauge bosons and free dual quarks. However,
there are also many differences. The major difference is that in
addition to the IR free gauge group there is another $SU$ gauge
group in the deconfined theory, which does not go free, and which
in its matter content itself has an antisymmetric tensor just like
the electric theory.  If one assumes that as argued above the $Sp$
group goes free, one can simply apply $a$-maximization to the
remaining $SU$ group to determine its evolution.


\begin{figure}[ht]
\centerline{\includegraphics[width=0.65\hsize]{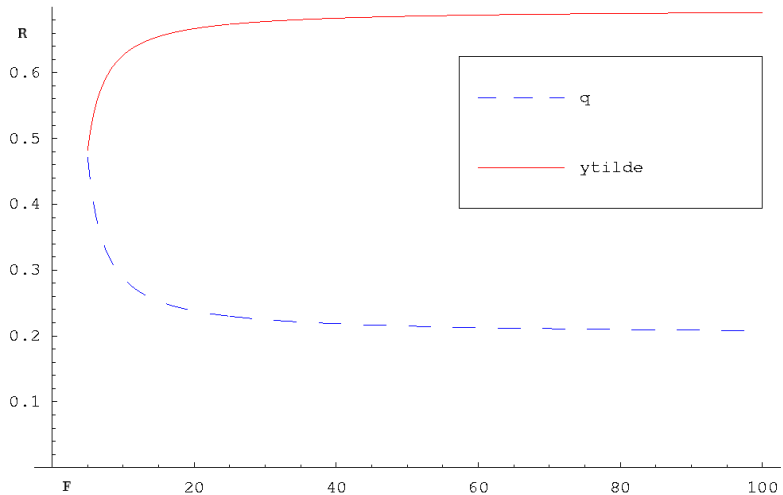}}
\caption{The $R$-charges of the independent fields are plotted as
a function of $F$.  The short dashed line represents the
$R$-charge of $q$, and the long dashed line represents
$\tilde{y}$. The smallest value that $F$ can be is $F=5$, for
$F<5$ this theory confines.} \label{rchargesu}
\end{figure}

The results from $a$-maximization of the remaining $SU$ group are
shown in Fig.~\ref{rchargesu}, where we plot the $R$-charges of
the independent fields $q$,$\tilde{y}$ as a function of the number
of flavors, $F$.  Note, that the size of the original gauge group
$N$ no longer affects the dynamics of the $SU$ group, which is now
only a function of $F$, this is why we can present the results as
a function of $F$ only.  However, it is still implicitly assumed
that $N$ is large enough for this mixed phase to occur at all
because for $N<8$ the $H$ meson will not go free and thus there
will not be a mixed phase.  What we find is that none of the gauge
invariants in the remaining $SU$ group go free as we reduce the
number of flavors before it confines at $F=4$. What this suggests
is that the mixed phase continues to exist down to $F=5$ and once
$F<5$ the theory confines which signals the end of the mixed
phase.


The mixed phase that occurs in this theory after $H$ decouples is
different than the normal picture that Seiberg duality shows for
other theories. For instance in an $SU(N)$ gauge theory with $F$
flavors Seiberg duality shows that below $F=3/2 N$ the theory goes
from a non-abelian Coulomb phase to a free magnetic phase. In the
case of the anti-symmetric tensor we have a product gauge group
dual and only one of the gauge groups goes free.

\section{Conclusion}
\label{conclusion} \setcounter{equation}{0}

The $SU(N)$ supersymmetric gauge theory with a two-index
antisymmetric tensor has many interesting properties not seen in
other theories.  There is no simple dual that unambiguously
defines the phases of this gauge theory but combining
``deconfinement" and the process of $a$-maximization we have
determined the phase evolution of this theory in terms of $N$ and
$F$.  This theory has a Banks-Zaks fixed point in the large
$N$,$F$ limit at $x=1/2$.  In the IR we believe that the theory is
in an interacting non-abelian Coulomb phase near the Banks-Zaks
fixed point.  As one increases $x$ the meson $M=Q\overline{Q}$
becomes a free field and decouples from the theory at $x=x_M\sim
2.95367$. Further increasing $x$ will cause the other meson in the
theory $H=\overline{Q} A\overline{Q}$ to become free at $x=x_H\sim
4.08952$.  At the point where the meson $H$ becomes free in the
deconfined dual description it dictates that one of the  dual
gauge groups, $Sp(2F-8)$, becomes free.  Therefore for $x>x_H$ the
electric theory ceases to be a good description of the physics and
one should use the dual deconfined description.  In this
deconfined description a mixed phase of the gauge theory continues
to exist where the $Sp(2F-8)$ group remains free and the $SU(F-3)$
group is in an interacting non-abelian Coulomb phase.  This mixed
phase exists for all $F \ge 5$; at the point where $F<5$ the
theory confines and the dynamics is well understood.  We expect
similar results for $SU(N)$ with a two-index symmetric tensor in
the large $N$,$F$ limit, except the free gauge group will be
$SO(2F+8)$.  While $a$-maximization does not a priori tell us the
phase structure of a theory, by combining it with the tools of
deconfinement and duality we have shown that a new kind of fixed
point can be reached.

\section*{Acknowledgments}
We thank Josh Erlich and Ken Intriligator for discussions and useful comments on the
manuscript. The research of C.C. and P.M. is supported
in part by the
NSF under grants PHY-0139738 and PHY-0098631, and in part by the DOE OJI grant
DE-FG02-01ER41206. The research of J.T. is supported by the US
Department of Energy under contract W-7405-ENG-36.

\end{document}